\documentclass{czjphys}         
\usepackage{graphicx}
\usepackage{epsf,psfrag,here}

\def\al{\!\!&\!\!}
\def\be{\begin{equation}}
\def\ee{\end{equation}}
\def\bea{\begin{eqnarray}}
\def\eea{\end{eqnarray}}

\newcommand{\bdm}{\begin{displaymath}}
\newcommand{\edm}{\end{displaymath}}
\newcommand{\no}{\nonumber \\}

\newcommand{\fs}{\; .}
\newcommand{\co}{\; ,}
\newcommand{\eff}{{e\hspace{-0.1em}f\hspace{-0.18em}f}}
\newcommand{\QCD}{{\mbox{\tiny Q\hspace{-0.05em}CD}}}
\newcommand{\indV}{{\scriptscriptstyle V}}

\newcommand{\indL}{{\scriptscriptstyle L}}
\newcommand{\indR}{{\scriptscriptstyle R}}
\newcommand{\lvac}{\langle 0|\,}
\newcommand{\rvac}{\,|0\rangle}
\renewcommand{\nc}{N_{\!c}}
\newcommand{\qbar}{\overline{\rule[0.42em]{0.4em}{0em}}\hspace{-0.45em}q}
\newcommand{\ubar}{\overline{\rule[0.42em]{0.4em}{0em}}\hspace{-0.5em}u}
\newcommand{\dbar}{\,\overline{\rule[0.65em]{0.4em}{0em}}\hspace{-0.6em}d}
\newcommand{\sbar}{\overline{\rule[0.42em]{0.4em}{0em}}\hspace{-0.5em}s}

\newcommand{\ind}{\scriptscriptstyle}
\newcommand{\R}{{\mbox{\tiny R}}}
\renewcommand{\L}{{\mbox{\tiny L}}}

\newcommand{\lbar}{\bar{\ell}}
\newcommand{\rs}{\langle r^2\rangle\rule[-0.2em]{0em}{0em}_s}
\newcommand{\lsim}{\,\raisebox{-0.3em}{$\stackrel{\raisebox{-0.1em}{$<$}}{\sim}$}\,}

\begin{document}
\title{Strong Interactions at Low Energies}
\authori{H.~Leutwyler}  
\addressi{Institute for Theoretical Physics, University of Bern\\
Sidlerstr.~5, CH-3012 Bern, Switzerland\\E-mail: leutwyler@itp.unibe.ch}
\authorii{}
\addressii{}
\authoriii{}    \addressiii{}
\authoriv{}     \addressiv{}
\authorv{}      \addressv{}
\authorvi{}     \addressvi{}
\headauthor{H.~Leutwyler}   
\headtitle{Strong Interactions at Low Energy}
\lastevenhead{H.~Leutwyler, Strong Interactions at Low Energy}
\refnum{A}
\daterec{3 Nov 2001}    
\issuenumber{0}  \year{2001}
\setcounter{page}{1}
\maketitle
\begin{abstract}
The lectures review some of the basic concepts relevant for an understanding
of the low energy properties of the strong interactions: chiral  
symmetry, spontaneous symmetry breakdown,  Goldstone bosons, quark condensate.
The effective field theory used to analyze the low energy structure is 
briefly sketched. As an illustration, I discuss the implications of
the recent data on the decay $K\rightarrow \pi\pi e\nu$ for the magnitude
of the quark condensate.
\begin{center} Lectures given at the $14^{th}$ Indian Summer School, Prague,
  July 2001.\end{center}
\end{abstract}

\section{Introduction}

Let us first locate the topic discussed in these lectures within the
Standard Model. In that framework, the dynamical variables are the gauge 
bosons $\gamma, 
W,Z,G$, the Higgs fields $(\phi_1,\phi_2,\phi_3,\phi_4)$, the
quarks $q$ and the leptons $\ell$. Except for the mass term of the
Higgs field, the Lagrangian does not contain mass parameters -- the masses of
the various particles are of dynamical origin:
The ground state contains a condensate of neutral Higgs particles,
$\lvac \phi_1\rvac\neq 0$. Neither the
photon nor the gluons take notice -- for these, the vacuum is transparent, 
because $\phi_1$ is electrically neutral and does not carry colour. 
For the gauge
fields that mediate the weak interaction, however, this is not the case:
The vacuum is not transparent for $W$ and $Z$ waves of low frequency -- these
particles do interact with those forming the condensate, because $\phi_1$ is 
not neutral 
with respect to flavour. As a consequence, the frequency of the  
$W$ and $Z$ waves tends to a nonzero value at large wavelength: The
corresponding 
particles move at a speed that is smaller than the velocity of light -- both 
the $W$ and the $Z$ pick up a mass. 

The quarks and leptons also interact 
with the particles in the condensate and thus also pick up mass.  
It so happens that the interactions of $\nu,e,\mu,u,d,s$ with the Higgs fields
are weak, so that the masses $m_\nu,m_e,m_\mu,m_u,m_d,m_s$ are small. 
The remaining fermion masses, as well as $m_W$, $m_Z$ and $m_H$ 
are not small. We do not know why the observed mass pattern looks like this,
but we can analyze the consequences of this empirical fact. 

At energies that are small compared to $\{m_W,m_Z,m_H\}=O(100\,\mbox{GeV})$, 
the weak interaction freezes out, because these
energies do not suffice to bridge the mass gap and to excite the corresponding
degrees of freedom.  
As a consequence, the gauge group of the Standard Model, 
SU(3)$\times$SU(2)$\times$U(1), breaks down to
the subgroup SU(3)$\times$U(1) -- only the photons, the gluons, the quarks and
the charged leptons are active at low energies. Since the neutrini
neither carry colour nor charge, they decouple. 

\section{Effective theory for {\boldmath
    $E\ll 100\,\mbox{GeV}$\unboldmath}} 

The Lagrangian relevant in the low energy domain is the one of QCD $+$ QED,
which is characterized by the two
coupling constants $g$ and $e$.  
In contrast to the Standard Model,
the SU(3)$\times$U(1) Lagrangian does contain mass terms: the quark and lepton
mass matrices $m_q$, $m_\ell$. 
Moreover, Lorentz and gauge invariance
permit the occurrence of a term 
proportional to the operator
\bea \omega=
\frac{1}{16\,\pi^2}\,\mbox{tr}\hspace{-0.6em}
\rule[-0.5em]{0em}{0em}_c\hspace{0.5em} G_{\mu\nu}\,\tilde{G}^{\mu\nu}\fs
\nonumber\eea
The corresponding coupling constant $\theta$ is referred to as 
the vacuum angle. 
The field basis may be chosen such that $m_q$ and $m_\ell$ are
diagonal and positive.
The fact that the electric dipole moment of the
neutron is very small implies that -- in this basis -- 
$\theta$ must be tiny. This is called the strong CP-problem: We do not
really understand why the 
neutron dipole moment is so small.

The two gauge fields involved in the effective low energy theory behave in a
qualitatively different manner: While the photons do not carry electric
charge, the gluons do carry colour. This difference is responsible for the fact
that the strong interaction becomes strong at low energies, while the
electromagnetic interaction becomes weak there, in fact remarkably weak:
The photons and leptons essentially decouple from the quarks and gluons.
The electromagnetic interaction can be accounted for by means of the
perturbation series in powers of $e$. For the QCD part of the theory, on the
other hand, perturbation theory is useful only at high energies. In the low
energy domain, the strong interaction is so strong that it 
confines the quarks and gluons.

The resulting effective low energy theory is mathematically more 
satisfactory than the 
Standard Model as such -- it does not involve scalar degrees of freedom and
has fewer free parameters. Remarkably, this simple theory 
must describe the
structure of cold matter to a very high degree of precision, once 
the parameters in the Lagrangian are known. It in particular 
explains the size of the atoms in terms of the scale
\bea a_{\hspace{-0.07em}\ind B}=\frac{4\,\pi}{e^2\,m_e}\co\nonumber\eea 
which only contains the two parameters $e$ and $m_e$ -- these are indeed known
to an incredible precision. Unfortunately, our ability to solve the QCD part
of the theory is rather limited -- in particular, we are
still far from being able to demonstrate on the basis of the QCD Lagrangian
that the strong interaction actually
confines colour. Likewise, our knowledge of the magnitude of the light quark
masses is still rather limited -- we need to know these more accurately 
in order to test ideas that might lead to an understanding of the
mass pattern, such as the relations with the lepton masses that emerge from 
attempts at unifying the electroweak and strong forces.

\section{Massless QCD -- a theoretical paradise}
In the following, I focus on the QCD part and switch the electromagnetic
interaction off.  As mentioned already, $m_u,m_d$ and $m_s$ happen to be
small. Let me first set these parameters equal to zero and, moreover, 
send the masses of the heavy quarks,
$m_c,m_b,m_t$ to infinity. In this limit, the theory becomes a theoreticians
paradise: The Lagrangian contains a single parameter, $g$. In fact, since 
the value of $g$ depends on the running scale used, the theory does not
contain any dimensionless parameter that would need to be adjusted to
observation. In principle, this theory fully specifies all dimensionless 
observables as pure numbers, while dimensionful quantities like masses or
cross sections can unambiguously
be predicted in terms of the scale $\Lambda_{\QCD}$ or the mass of the proton.
The resulting theory -- QCD with three massless flavours -- is among the
most beautiful quantum field theories we have. I find it breathtaking that, 
at low energies, nature reduces to this beauty, as soon as the dressing with
the electromagnetic interaction is removed and the Higgs condensate is 
replaced by one that does not hinder the light quarks, but is impenetrable for
$W$ and $Z$ waves as well as for heavy quarks.

The Lagrangian of the massless theory, which I denote by ${\cal L}_{\QCD}^0$,
has a high degree of symmetry, which originates in the fact that the
interaction among the quarks and gluons is flavour-independent and conserves
helicity: ${\cal L}_{\QCD}^0$ is invariant under independent flavour rotations
of the three right- and left-handed quark fields. These form the group
$G=\mbox{SU(3)}_\R\times\mbox{SU(3)}_\L$. The corresponding 16 currents
$V^\mu_i\qbar\gamma^\mu\frac{1}{2}\,\lambda_i q$ and 
$A^\mu_i=\qbar\gamma^\mu\gamma_5\frac{1}{2}\,\lambda_i q$ are conserved, so
that their charges commute with the Hamiltonian:
\bea [\,Q_i^{\mbox{\tiny V}},H_\QCD^0\,]=
[\,Q_i^{\mbox{\tiny A}},H_\QCD^0\,]=0\co
\hspace{2em}i=1,\,\ldots\,,8\fs\nonumber\eea
Vafa and Witten~\cite{Vafa Witten} have shown that the state of lowest energy
is necessarily invariant under the vector charges: 
$Q_i^{\mbox{\tiny V}}\rvac=0$. For the axial charges, however, there are the
two possibilities characterized in table 1.
\begin{table}
\begin{tabular}{|c|c|}
\hline
\rule[-0.7em]{0em}{1.9em}$Q_i^{\mbox{\tiny A}}\rvac=0$&
$Q_i^{\mbox{\tiny A}}\rvac\neq0$\\ \hline\rule{0em}{1.2em}
Wigner-Weyl realization of $G$&Nambu-Goldstone realization of $G$ \\
ground state is symmetric & ground state is asymmetric\\
\rule[-1em]{0em}{2em}
$\lvac\qbar_\R q_\L\rvac = 0$&$\lvac\qbar_\R q_\L\rvac \neq 0$
\vspace*{-0.5em}\\
ordinary symmetry & spontaneously broken symmetry\\
spectrum contains parity partners & spectrum contains Goldstone bosons\\
\rule[-0.7em]{0em}{0em}degenerate multiplets of $G$& 
degenerate multiplets of $\mbox{SU(3)}\in G$\\
\hline\end{tabular}
\caption{Alternative realizations of the symmetry group
  $G=\mbox{SU(3)}_\R\times\mbox{SU(3)}_\L$.} 

\vspace*{-1em}
\end{table}

The observed spectrum does not contain parity doublets. In the case
of the lightest meson, the $\pi(140)$, for instance, the lowest state with 
the same spin and flavour quantum numbers, but opposite parity is the 
$a_0(980)$. So, expe\-ri\-ment rules out the first possibility. 
In other words, 
for dynamical reasons that yet remain to be understood,
the state of lowest energy is an asym\-metric state. 
Since the axial charges
commute with the Hamiltonian, there must be eigenstates with the same energy
as the ground state:
\bea H^0_\QCD\, Q_i^{\mbox{\tiny A}}\rvac= Q_i^{\mbox{\tiny A}}\,H^0_\QCD\rvac
=0\fs\nonumber\eea
The spectrum must contain 8 states $Q_1^{\mbox{\tiny A}}\rvac,\ldots\,,
Q_8^{\mbox{\tiny A}}\rvac$  
with $E=\vec{P}=0$, describing massless particles, the Goldstone bosons of the
spontaneously broken symmetry. Moreover, these must carry spin 0, negative
parity and form an octet of SU(3). 
 
\section{Spontaneous and explicit symmetry breaking}
Indeed, the 8 lightest hadrons,
$\pi^+,\pi^0,\pi^-,K^+,K^0,\bar{K}^0,K^-,\eta$,
do have these quantum numbers, but massless they
are not.
This has to do with the deplorable fact that we are not living in paradise: the
masses $m_u,m_d,m_s$ are different from zero and thus allow the left-handed
quarks to communicate with the right-handed ones. 
The full Hamilitonian is of the form
\bea  H_\QCD\al=\al H^{\ind 0}_\QCD+ H^{\ind 1}_\QCD\co\no 
H^{\ind 1}_\QCD\al=\al\int\!\!d^3\!x\;
\qbar_\indR m\, q_\indL+\qbar_\indL m^\dagger q_\indR\co\hspace{2em}
m=\left(\!\!\!\mbox{\begin{tabular}{ccc}\vspace*{-0.5em}$m_u\!\!\!$&
&\\\vspace*{-0.5em}&$\!\!\!m_d\!\!\!
$&\\&&$\!\!\!m_s$
\end{tabular}}\!\!\!\!\right)\fs\nonumber\eea
The quark masses may be viewed as symmetry breaking parameters: the 
QCD-Hamiltonian is only approximately symmetric under independent rotations of
the right- and left-handed quark fields, to the extent that these parameters
are small. Chiral symmetry is thus broken in two ways: 
\begin{itemize}\item spontaneously\hspace{3em} 
$\lvac\qbar_\indR q_\indL\rvac \neq 0$
\item explicitly\hspace{5.2em} $m_u,m_d,m_s\neq 0$\end{itemize}

Since the masses of the two lightest quarks are particularly small, the
Hamiltonian of QCD is almost exactly invariant under the subgroup 
SU(2)$_{\indR}\times$SU(2)$_{\indL}$. The ground state
spontaneously breaks that symmetry to the subgroup SU(2)$_{\indV}$ --
the good old isospin symmetry discovered in the thirties of the last
century \cite{Heisenberg}. The pions represent the
corresponding Goldstone bosons \cite{Nambu}, while the kaons and 
the $\eta$ remain
massive if the limit $m_u,m_d\rightarrow 0$ is taken at fixed $m_s$. 
In the following, I consider this framework and, moreover, ignore isospin
breaking, setting $m_u=m_d=\hat{m}$.

If SU(2)$_{\indR}\times$SU(2)$_{\indL}$ was an exact symmetry, the pions 
would be strictly massless. According to
Gell-Mann, Oakes and Renner \cite{GMOR}, the square of the pion mass is
proportional to the product of the quark masses and the quark condensate:
\bea\label{eq:GMOR} 
\al\al M^2_\pi\simeq\frac{1}{F_\pi^2}\times(m_u+m_d)\times |\lvac\, \ubar 
u \rvac|\fs\eea
The factor of proportionality is given by the pion decay constant
$F_\pi$. The term $m_u+m_d$ measures the explicit
breaking of chiral symmetry,  
while the quark condensate, 
\bdm \lvac\,\ubar u\rvac =\lvac\, \ubar_{\indR} 
u_{\ind L}\rvac +\mbox{c.c.}=
\lvac\,\dbar d\rvac\co\edm
is a measure of the spontaneous symmetry breaking: it may be viewed as an
order parameter and plays a role analogous to the spontaneous 
magnetization of a magnet.

The approximate
validity of the relation (\ref{eq:GMOR}) was put to question by Stern and
collaborators \cite{KMSF}, who pointed out that there is no experimental
evidence for the quark condensate to be different from zero. Indeed,
the dynamics of the ground state of QCD is not understood -- it could resemble
the one of an antiferromagnet, where, for dynamical reasons, the
most natural candidate for an order parameter, the magnetization, 
happens to vanish. There are a number of theoretical reasons indicating 
that this scenario is unlikely:

(i) The fact that the pseudoscalar meson octet satisfies the Gell-Mann-Okubo 
formula remarkably well would then be accidental.

(ii) The value obtained for the quark condensate on the basis of QCD sum
rules, in particular for the baryonic correlation functions \cite{Ioffe}, 
confirms the standard picture.

(iii) The lattice values \cite{Lubicz} for the ratio $m_s/\hat{m}$
agree very well with the result of 
the standard chiral perturbation theory analysis
\cite{Leutwyler 1996}, corroborating this picture further.

Quite irrespective, however, of whether or not the scenario advocated 
by Stern et al.~is theoretically appealing, the issue can be subject to 
experimental test. In fact, significant progress has recently been achieved 
in this direction \cite{CGLPRL,Pislak}. 

\section{Chiral perturbation theory}
\label{eff}
The consequences of the fact that the explicit symmetry breaking is small may
be worked out by means of an effective field theory, ``chiral perturbation
theory'' \cite{Weinberg Physica} --\cite{reviews}. In this context, 
the heavy quarks do not play an important role -- 
as the corresponding fields are singlets under 
chiral transformations of the light flavours, we may include their 
contributions in the symmetric part of the Hamiltonian, 
irrespective of the size of their mass. 

Concerning the strange
quark, there are two options: we may either treat the corresponding mass
term $m_s \sbar s$ as a perturbation, so that the unperturbed Hamiltonian
$H^{\ind 0}_\QCD$ is invariant under SU(3)$_{\indR}\times$SU(3)$_{\indL}$.
Alternatively, we may treat the strange quark on the same footing as the heavy 
ones, including its mass term in $H^{\ind 0}_\QCD$. The symmetry group
then reduces to SU(2)$_{\indR}\times$SU(2)$_{\indL}$ and the spontaneous 
symmetry breakdown gives rise to only 3 Goldstone bosons, the pions.
The effective theories are different in the two cases. In the following,
I consider the second option.

At low energies, the behaviour of scattering amplitudes or current matrix
elements can be described in terms of a {\it Taylor series expansion} in powers
of the momenta.
The electromagnetic form factor of the pion, e.g., may be
exanded in powers of the momentum transfer $t$.
In this case, the first two Taylor coefficients are related to the total charge
of the particle and to the mean square radius of the charge distribution,
respectively,
\be \label{taylor}
f_{\pi^+}(t) = 1 + \mbox{$\frac{1}{6}$} \langle r^2\rangle_{\pi^+}\, t +
O(t^2)\fs \end{equation}
Scattering lengths and effective ranges are analogous low energy
constants occurring in the Taylor series expansion of scattering amplitudes.

The occurrence of light particles gives rise to singularities in the low
energy domain, which limit the range of validity of the Taylor series
representation. The form factor $f_{\pi^+}(t)$, e.g., contains a branch cut
at $t=4 M_\pi^2$, such that the formula (\ref{taylor}) provides an adequate
representation only for $|t|\ll 4 M_\pi^2$. The problem becomes even more
acute if $m_u$ and $m_d$ are set equal to zero. The pion mass then
disappears, the branch cut sits at $t=0$ and the Taylor series does not
work at all. I first discuss the method used in the low energy analysis for
this extreme case, returning
to the physical situation with $m_u,m_d\neq 0$ below.

The reason why the spectrum of QCD with two massless quarks contains three
massless bound states is understood:
they are the Goldstone bosons of a
hidden symmetry. The symmetry, which
gives birth to these, at the same time also determines their low energy
properties. This makes it possible to explicitly work out
the poles and branch cuts generated by the exchange of Goldstone bosons.
The remaining singularities are
located comparatively far from the origin, the nearest one being due to
the $\rho$-meson. The result is a modified Taylor series expansion in powers
of the momenta, which works, despite the presence of massless particles.
In the case of the $\pi\pi$ scattering amplitude,
e.g., the radius of convergence of the modified series
is given by $s=M_\rho^2$, where $s$ is the square of the energy in the center
of mass system (the first few
terms of the series only yield a decent description of the amplitude if
$s$ is smaller than the radius of
convergence, say $s\!<\!\frac{1}{2}M_\rho^2\rightarrow \sqrt{s}\!<
540\;\mbox{MeV}$).

As pointed out by Weinberg \cite{Weinberg Physica}, the modified expansion
may explicitly be constructed by means of an effective field theory, which is
referred to as {\it chiral perturbation theory} and involves the following
ingredients: \\ (i) The quark and gluon fields of QCD are
replaced by a set of pion fields, describing the degrees of freedom of the
Goldstone
bosons. It is convenient to collect these in a
$2\!\times\!2$
matrix U$(x)\!\in\,$SU(2). \\
(ii) The Lagrangian of QCD is replaced by an
effective Lagrangian, which only involves the field U$(x)$, and its derivatives
\bdm {\cal L}_{\QCD}\;\;\longrightarrow \;\;{\cal L}_\eff(U,\partial
U,\partial^2U,\ldots)\fs\edm
(iii) The low energy expansion corresponds to an
expansion of the effective
Lagrangian, ordered according to the number of the derivatives of the field
$U(x)$.
Lorentz invariance only permits terms with an even number
of derivatives,
\bdm
{\cal L}_{\eff}= {\cal L}_{\eff}^{\,0} + {\cal L}_{\eff}^{\,2} +
{\cal L}_{\eff}^{\,4} + {\cal L}_{\eff}^{\,6} +
\ldots
\edm

Chiral symmetry very strongly constrains the form of the terms occurring
in the series. In particular, it excludes momentum
independent interaction vertices:
Goldstone bosons can only interact if they carry momentum. This property
is essential for the consistency of the low energy analysis, which treats the
momenta as expansion parameters. In the notation introduced above,
chiral symmetry implies that the leading term ${\cal L}_{\eff}^{\,0}$ is an
uninteresting constant -- up to a sign, this term represents the
vacuum energy of QCD in the chiral limit. The first nontrivial contribution 
involves 
two derivatives,
\be \label{eff1}
{\cal L}_{\eff}^{\,2} = \mbox{$\frac{1}{4}$}F^2 \mbox{tr} \{
\partial_\mu U^+ \partial^\mu U \} \co
\ee
and is fully determined by the pion decay constant. At order $p^4$, the
symmetry permits two independent terms,\footnote{In the framework of the
effective theory, the anomalies of QCD
manifest themselves through an extra contribution,
the Wess-Zumino term, which is also of order $p^4$ and is proportional to the
number of colours.}
\be\label{eff3} {\cal L}_{\eff}^{\,4}=\mbox{$\frac{1}{4}$}l_1 (\mbox{tr} \{
\partial_\mu U^+ \partial^\mu U \})^2
+ \mbox{$\frac{1}{4}$}l_2\mbox{tr} \{
\partial_\mu U^+ \partial_\nu U \}\mbox{tr} \{
\partial^\mu U^+ \partial^\nu U \}\co\ee
etc. For most applications, the derivative expansion
is needed only to this order.

The most remarkable property of the method is that it does not
mutilate the theory under investigation:
The effective field theory framework is no more
than an efficient machinery, which
allows one to work out the modified Taylor series, referred to above.
If the effective Lagrangian includes all of the terms
permitted by the symmetry, the
effective theory is mathematically equivalent to QCD
\cite{Weinberg Physica,found}.
It exclusively exploits the symmetry properties of QCD and involves an infinite
number of effective coupling constants,
$F,l_1,l_2,\ldots\;$, which represent the Taylor coefficients of the
modified expansion.

In QCD, the
symmetry, which controls the low energy properties of the Goldstone bosons, is
only an approximate one. The constraints imposed by the hidden,
approximate symmetry can still be worked out, at the price of
expanding the
quantities of physical interest in powers of the symmetry breaking parameters
$m_u$ and $m_d$. The low energy analysis then involves a combined
expansion,
which treats both, the momenta and the quark masses as small parameters.
The effective Lagrangian picks up additional terms, proportional to powers of
the quark mass matrix,
\bdm
m = \left(\mbox{\raisebox{0.4em}{$ m_u $}}\;\mbox{\raisebox{-0.4em}
{$m_d$}}\,
\right)
\edm
It is convenient to count $m$ like two powers of
momentum, such that the expansion of the effective Lagrangian still starts at
$O(p^2)$ and only contains even
terms. The leading contribution picks up a term linear in $m$,
\be\label{eff2}
{\cal L}_{\eff}^{\,2} = \mbox{$\frac{1}{4}$}F^2 \mbox{tr} \{
\partial_\mu U^+ \partial^\mu U \} +\mbox{$\frac{1}{2}$}F^2B\,\mbox{tr}
\{m(U+U^\dagger)\}\fs
\ee
Likewise, ${\cal L}_\eff^4$ receives additional contributions, involving
further effective coupling constants, etc. 

\section{Derivation of the Gell-Mann-Oakes-Renner relation}
The expression (\ref{eff2}) represents a compact summary of the soft pion
theorems
established in the 1960's: The leading terms in the low energy expansion of the
scattering amplitudes and current matrix elements are given by the tree graphs
of this Lagrangian. 

I illustrate the content of this statement with a derivation of the 
Gell-Mann-Oakes-Renner 
relation on the basis of the effective theory. For this purpose, we 
first need to evaluate the pion mass as well as the quark condensate and the 
pion decay constant at tree level and then to verify that the three quantities
are indeed related according to eq.~(\ref{eq:GMOR}). The
tree graphs of the Lagrangian (\ref{eff2}) represent the corresponding
classical action. To work out this action, it is convenient to
introduce a set of coordinates on SU(2), so that the matrix field $U(x)$ is 
expressed in terms of a set of three ordinary scalar fields 
$\vec{\pi}=\pi^1(x),\pi^2(x),\pi^3(x)$ 
that describe the three different pion flavours. The coordinates
are a matter of choice. Most of the calculations are performed in canonical 
coordinates, which are defined by
\bea\label{eq:canonical} U=\exp \frac{i\,\vec{\pi}\cdot\vec{\tau}}{F}\fs\eea
An equally suitable choice is
\bea\label{eq:sigma} U=\sqrt{1-\frac{\vec{\pi}^2}{F^2}}+
\frac{i\,\vec{\pi}\cdot\vec{\tau}}{F}\fs\eea
The results obtained from the effective theory are coordinate independent.
Quantities of physical interest are obtained by evaluating the path integral 
over the effective Lagrangian -- the pion field merely
represents the variable of integration in that integral and
does not have physical significance. In particular,
in tree graph approximation, only the value of the classical action
at the extremum is relevant -- this value evidently is independent of 
the manner in which the matrix $U(x)$ is parametrized. In particular,
the factors of $1/F$ occurring in the above formulae could just as well 
be dropped -- the parametrization $U=\exp i\,\vec{\pi}\cdot \vec{\tau}$, 
for instance, which involves a dimensionless pion field, is equally 
legitimate. 

Up to and including terms quadratic in the pion field, the two 
parametrizations introduced above yield the same expression for the 
effective Lagrangian:
\bea\label{eq:Lquad} {\cal L}_{\eff}^{\,2} =(m_u+m_d)F^2 B+ 
\mbox{$\frac{1}{2}$}\,\partial_\mu\vec{\pi}\,\partial^\mu\vec{\pi}-
\mbox{$\frac{1}{2}$}\,(m_u+m_d) B\,\vec{\pi}^{\,2}+O(\vec{\pi}^{\,4})\fs\eea
Remarkably, only the sum $m_u+m_d$ of the quark masses matters. Indeed,
it is easy to show that this is true also of the contributions involving four 
or more powers of the pion field and which describe the interaction among the 
Goldstone bosons: if $U$ is a unitary $2\times 2$ matrix with $\det U=1$,
then $U+U^\dagger$ is proportional to the unit matrix. Accordingly, the
Lagrangian (\ref{eff2}) only involves the trace of the quark mass matrix,
that is the sum  $m_u+m_d$. In other words, the pions are protected from
the isospin breaking effects generated by the mass difference $m_u-m_d$: 
the leading order expression for the effective Lagrangian is isospin 
invariant. 

The third term immediately shows that the square of the pion mass 
grows linearly with the quark masses:
\bea M_\pi^2=(m_u+m_d)B+O(m^2)\fs\nonumber\eea
Note that the tree graph approximation of the effective theory only yields
the leading term in the expansion of $M_\pi^2$ in powers of the quark 
masses. I will discuss the corrections of order $m^2$ in detail below.

The first term in eq.(\ref{eq:Lquad})
represents a contribution to the vacuum energy. It shows that,
in tree graph approximation of the
effective theory, the vacuum picks up an energy shift 
proportional to the quark masses: 
\bdm \Delta E_0=-V(m_u+m_d)F^2 B\co\edm
where $V$ is the space volume. This expression must agree with the energy
shift evaluated within the underlying theory. To leading order in the quark 
masses, the energy shift is given by the expectation value of the
perturbation:
\bdm \Delta E_0=\lvac H^{\ind 1}_\QCD\rvac=
V \lvac m_u\, \ubar\, u +m_d \,\dbar\, d\rvac\fs\edm 
For the tree graphs of the
effective theory to reproduce this expression, we must have
\bdm \lvac m_u\, \ubar\, u +m_d \,\dbar\, d\rvac=-(m_u+m_d)F^2 B+O(m^2)\fs\edm
Stated otherwise, the value of the quark condensate in the chiral limit
is determined by the effective coupling constants $F$ and $B$: 
\bdm \lvac \ubar u\rvac\,\rule[-0.53em]{0.04em}{1.25em}_{\,m_u,m_d
\rightarrow 0} =\lvac \dbar d\rvac\,\rule[-0.53em]{0.04em}{1.25em}_{\,m_u,m_d
\rightarrow 0}=-F^2 B\fs\edm

Putting the results for $M_\pi^2$ and for the quark condensate together,
we indeed arrive at the Gell-Mann-Oakes-Renner formula (\ref{eq:GMOR}),
except for one point: $F_\pi$ is replaced by $F$. Indeed, the coupling
constant $F$ represents the leading term in the 
expansion of the pion decay constant\footnote{In the normalization used here, 
the experimental value is $F_\pi=92.4$ MeV.} in powers of $m_u,m_d$ 
\bea\label{eq:Ftree} F_\pi=F+O(m)\fs\nonumber\eea 
To verify the validity of this formula, we need to work out the 
vacuum-to-pion matrix element of
the axial current. In the framework of the effective theory, the 
representation for the axial current may be obtained by applying the 
Noether theorem
to the above effective Lagrangian. The result reads
\bdm \vec{A}_\mu=-F\,\partial_\mu \vec{\pi}+O(\vec{\pi}^{\,3})\fs\edm
This shows that, in tree approximation of the effective theory, we indeed
have
\bdm \lvac A^i_\mu|\pi\rangle=i\,p_\mu\,F\co\edm
in accordance with the above claim.

\section{Chiral expansion  beyond leading order} 
The effective field theory
represents an efficient and systematic framework, which allows one to work out
the corrections to the soft pion predictions, those arising from the
quark masses as well as those from the terms of higher order
in the momenta. The evaluation is based on a perturbative
expansion of the quantum fluctuations of the effective field. In addition to
the tree graphs relevant for the soft pion results, graphs containing vertices
from the higher order contributions ${\cal L}_\eff^4,{\cal L}_\eff^6\ldots$ and
loop graphs
contribute. The leading term of the effective Lagrangian describes
a nonrenormalizable theory, the "nonlinear $\sigma$-model". The
higher order terms in the derivative expansion, however, automatically contain
the relevant counter terms. The divergences occurring in the loop
graphs merely renormalize the effective coupling constants. The effective
theory is a perfectly renormalizable scheme, order by order in the low
energy expansion, so that, in principle, the result of the calculation does not
depend on who it is who did it.

I now illustrate this with the terms occurring in the 
chiral expansion of $M_\pi$ and $F_\pi$ at next-to-leading order. 
For this purpose, we need to know the effective Lagrangian
to next-to-leading order. As mentioned in section \ref{eff},
the expression in eq.(\ref{eff3}) only holds in the chiral limit. 
The symmetry breaking generated by
the quark masses gives rise to additional contributions
that are linear or quadratic in $m$. Ignoring terms that are independent
of the pion field and thus only contribute to the vacuum energy, the full 
expression reads
\bea\label{eff4} {\cal L}_{\eff}^{\,4}\al=\al
\mbox{$\frac{1}{4}$}\,l_1 (\mbox{tr} \{
\partial_\mu U^+ \partial^\mu U \})^2
+ \mbox{$\frac{1}{4}$}\,l_2\mbox{tr} \{
\partial_\mu U^+ \partial_\nu U \}\mbox{tr} \{
\partial^\mu U^+ \partial^\nu U \}\\
\al\al+\mbox{$\frac{1}{4}$}\,l_4\,B\mbox{tr}
\{\partial_\mu U^+ \partial_\nu U \}\mbox{tr}\{m(U+U^\dagger)\}
+\mbox{$\frac{1}{4}$}(l_3+l_4)\,B^2(\mbox{tr}\{m(U+U^\dagger)\})^2
\no\al\al+\mbox{$\frac{1}{4}$}\,l_7\,B^2(\mbox{tr}\{m(U-U^\dagger)\})^2
\fs\nonumber\eea
The queer numbering of the coupling constants is connected with the fact
that we are not making use of external fields, so that 
$l_5$ and $l_6$ do not show up. External fields
represent a very useful tool for the formulation of the 
effective theory, but we do not need them here.

The last term describes isospin breaking effects: it vanishes if $m_u$ is set 
equal to $m_d$. To verify this
statement, we may note that, for $m_u=m_d$, the quark 
mass matrix is proportional to the unit matrix, so that the term becomes
proportional to the square of $\mbox{tr}(U-U^\dagger)$. The representation
(\ref{eq:sigma}) shows, however, that for $U\in \mbox{SU(2)}$, 
the quantity $\mbox{tr}\, U$ is real, so that 
$\mbox{tr}\,U=\mbox{tr}\,U^\dagger$. This shows that 
$(\mbox{tr}\{m(U-U^\dagger)\})^2$ vanishes in the limit $m_u=m_d$.

For the analysis of $F_\pi$ and $M_\pi$, we again only need the terms 
quadratic in the pion field. In the following, I work in the isospin symmetry
limit, setting
\bea m_u=m_d=m\fs\eea
To first nonleading order, the quadratic part of the effective Lagrangian
then takes the form
\bdm {\cal L}_{\eff}^{\,2}+ {\cal L}_{\eff}^{\,4}=
\frac{1}{2}\left\{1+\frac{4\, l_4\,m\, B}{F^2}\right\} 
\partial_\mu\vec{\pi}\, 
\partial^\mu\vec{\pi}-B\,m\left\{1+\frac{4\,B\,m\,(l_3+l_4)}{F^2}\right\}
\vec{\pi}^{\,2}+\ldots\edm
The structure of the terms is the same as for 
${\cal L}_{\eff}^{\,2}$ in eq.(\ref{eq:Lquad}), so that we can repeat the 
previous calculation, merely accounting
for the change in the coefficients. In this way, one easily checks that
the tree graphs of   
${\cal L}_{\eff}^{\,2}+{\cal L}_{\eff}^{\,4}$ yield
\bea \label{eq:MFtree}M_\pi^2=
2\,B\,m\left\{1+\frac{4\,B\,m\,l_3}{F^2}\right\}\co
 \hspace{2em}
F_\pi= F\left\{1+\frac{2\,B\,m\,l_4}{F^2}\right\}\fs\nonumber\eea

\section{Chiral logarithms}
At leading order, the tree graph approximation of the effective theory
provides the full answer. For
the effective theory to agree with QCD also at first nonleading
order of the expansion in powers of the quark masses, it does not suffice
to account for the tree graphs of ${\cal L}_{\eff}^{\,4}$, but we need
to also include the contributios from the one-loop graphs generated by the 
vertices of ${\cal L}_{\eff}^{\,2}$. For a general discussion of the power
counting that underlies this statement, I refer to the original papers 
\cite{Weinberg Physica,GL 1984}. In the case of the pion mass, a single one 
loop graph occurs, a tadpole, which is very easy to evaluate.
The graph generates a contribution proportional to the pion propagator at
the origin:
\bea M_\pi^2=M^2\left\{1+\frac{2\,M^2\,l_3}{F^2}-\frac{i\,\Delta(0)}{2\,F^2}
+O(M^4)\right\}\co\hspace{2em}M^2\equiv 2\,B\,m\fs\eea
In dimensional regularization, we have
\bea \Delta(0)=\frac{1}{(2\pi)^{d}}\!\!\int\! \frac{d^dp}{M^2-p^2-i\,\epsilon}=
\frac{i\,M^{d-2}\,\Gamma(1-\mbox{$\frac{d}{2}$})}{(4\pi)^{d/2}}\fs\nonumber\eea
The expression contains a pole at $d= 4$, with a residue that is proportional
to $M^2$:
\bea \Delta(0)\al=\al i\,M^2\left\{2\,\lambda+\frac{1}{16\pi^2}\ln 
\frac{M^2}{\mu^2}+O(d-4)\right\}\co\no
\lambda\al=\al \frac{\mu^{d-4}}{16\pi^2}\left\{\frac{1}{d-4}-
\frac{1}{2}(\ln 4 \pi+\Gamma'(1)+1)\right\}\fs\nonumber\eea
The divergence can thus be absorbed in
a renormalization of $l_3$,
\bdm l_r=l_3^r-\mbox{$\frac{1}{2}$}\lambda\co\edm
so that the result for $M_\pi^2$ stays finite, as it should:
\bea M_\pi^2=M^2\left\{1+\frac{2\,M^2\,l_3^r}{F^2}+\frac{M^2}{32 \pi^2 F^2}
\;\ln\frac{M^2}{\mu^2}+O(M^4)\right\}\fs\eea
Note that the
divergence is accompanied by a logarithm of $M$ -- the expansion of
$M_\pi^2$ in powers of the quark masses is not a simple Taylor series,
but contains a term of the type $M^4\ln M^2$ at first nonleading order.
In the above formula, the scale of this logarithm is the running
renormalization scale $\mu$ used in dimensional regularization. That scale
is arbitrary -- the running coupling constant $l_3^r$ depends on it in such
a manner that the expression for $M_\pi^2$ is independent of that scale.
The result may be written in the more transparent form
\bea\label{eq:Mexp} M_\pi^2=M^2-\frac{M^4}{32 \pi^2 F^2}
\;\ln\frac{\Lambda_3^2}{M^2}+O(M^6)\co\eea
where $\Lambda_3$ is the renormalization group invariant scale of $l_3$,
defined by
\bdm l_3^r=-\frac{1}{64\pi^2}\ln\frac{\Lambda_3^2}{\mu^2}\fs\edm

The symmetry does not determine the
numerical value of this scale. The crude estimates 
underlying the standard version of chiral perturbation theory \cite{GL 1984}
yield numbers in the range 
\bea \label{eq:Lambda3}0.2\;\mbox{GeV}<\Lambda_3<2\;\mbox{GeV}\fs\eea
The term of order $M^4$ is then very small compared to the one of order $M^2$,
so that the Gell-Mann-Oakes-Renner formula is obeyed very well.
Stern and collaborators investigate the more general framework, referred to as
``generalized chiral perturbation theory'', where
arbitrarily large values of $\lbar_3$ are considered. The
quartic term in eq.~(\ref{eq:Mexp}) can then take values comparable to the
``leading'', quadratic one. If so, the dependence of $M_\pi^2$ on the quark 
masses would fail to be approximately linear, even for values of $m_u$ and 
$m_d$ that are small compared to the intrinsic scale of QCD. A different 
bookkeeping for the terms 
occurring in the chiral perturbation series is then needed \cite{KMSF} 
-- the standard chiral power counting is adequate only if $\lbar_3$ is 
not too large.

The behaviour of the ratio $M_\pi^2/M^2$ as a function of $\hat{m}$ is
indicated in fig.~\ref{fig:FM}, taken from ref.~\cite{bangalore}. 
The fact that the information about the
value of $\Lambda_3$ is very meagre shows up through very large uncertainties.
In particular, 
with $\Lambda_3\simeq 0.5 \,\mbox{GeV}$, the ratio $M_\pi^2/M^2$ would remain
close to 1, on the entire interval shown. Note that outside the range
(\ref{eq:Lambda3}),
the dependence of $M_\pi^2$ on the 
quark masses would necessarily exhibit strong curvature. 

The figure illustrates the fact 
that brute force is not the only way the very small values of
$m_u$ and $m_d$ observed in nature can be reached through numerical
simulations on a  
lattice. It suffices to equip the strange quark with the physical value of
$m_s$ and to measure the dependence
of the pion mass on $m_u,m_d$ in the region where $M_\pi$ is comparable to
$M_K$. A fit to the data based on eq.(\ref{eq:Mexp})  
should provide an extra\-po\-lation to the 
physical quark masses that is under good control\footnote{The logarithmic 
singularities occurring at
next-to-next-to-leading order are also known
\cite{Colangelo 1995} -- for a
detailed discussion, I refer to \cite{opus2}.}. Moreover, the fit would
allow a determination of the scale $\Lambda_3$ on the lattice. This is of 
considerable interest, because that scale
also shows up in other contexts, in the $\pi\pi$ scattering lengths, 
for example. For recent work in this direction, I refer
to \cite{Heitger,Durr}. 

\section{Expansion of $\mathbf{F_\pi}$}
\begin{figure}[t] \centering
\psfrag{y}{}
\psfrag{Fpi}{\hspace{-3em}\raisebox{-0.4em}{$F_\pi/F$}}
\psfrag{Mpi}{\hspace{1em}\raisebox{0.6em}{$
M_\pi^2/M^2$}}
\psfrag{m}{\raisebox{-1.3em}{\hspace{-10em}$\hat{m}/m_s$}}

\mbox{\epsfysize=4.5cm \epsfbox{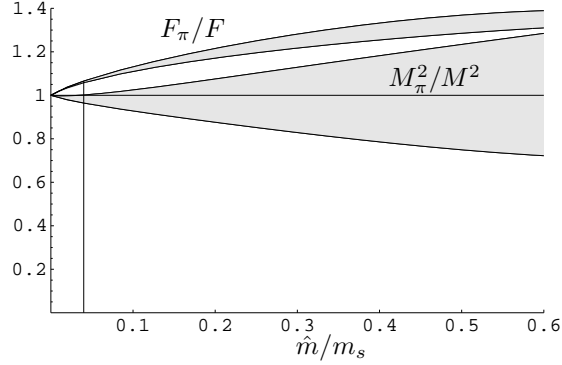} }\rule[-1em]{0em}{0em}

\caption{\label{fig:FM}Dependence of the ratios $F_\pi/F$ and $M_\pi^2/M^2$ on 
$\hat{m}=\frac{1}{2}(m_u+m_d)$. 
The strange quark mass is held fixed at the
physical value. The vertical line corresponds to the physical value of
$\hat{m}$.} 

\end{figure}

For the pion decay constant, the expansion analogous to eq.~(\ref{eq:Mexp}) 
reads
\bea \label{eq:Fexp}F_\pi = F\left\{1+\frac{\lbar_4\,M^2}{16\pi^2 F^2}
+O(M^4)\right\}\co\hspace{2em}\lbar_4=\ln\frac{\Lambda_4^2}{M^2}\fs\eea
In this case, the relevant effective coupling constant
is known rather 
well: chiral symmetry implies that it also 
determines the
slope of the scalar form factor of the pion,
\bea F_s(t)=
\langle\pi(p')|\;\ubar u + \dbar d\,|\pi(p)\rangle=F_s(0)\left\{1+
\mbox{$\frac{1}{6}$}\,\langle r^2\rangle_s\,t+O(t^2)\right\}\fs\nonumber\eea
As shown in ref.~\cite{GL 1984}, the expansion of $\langle r^2\rangle_s$ in 
powers of $m_u,m_d$ starts with
\bea\label{eq:rs}\langle r^2\rangle_s=\frac{6}{(4\pi F)^2}
\left\{\lbar_4-\frac{13}{12}+O(M^2)\right\}\fs\eea
Analyticity relates the scalar form factor to the $I=0$ $S$--wave phase shift
of $\pi\pi$ scattering \cite{DGL}. Evaluating the relevant dispersion relation
with the remarkably accurate information about the phase shift that
follows from the Roy equations \cite{opus2}, one finds
$\langle r^2\rangle_s=0.61\pm 0.04\,\mbox{fm}^2$. Expressed in terms of the
scale $\Lambda_4$, this amounts to 
\bea\label{eq:Lambda4} \Lambda_4=1.26\pm 0.14\;\mbox{GeV}\fs\eea 
Fig.~\ref{fig:FM} shows that this information determines the quark mass
dependence of the decay constant to within rather narrow limits. 
The change in $F_\pi$ occurring 
if $\hat{m}$ is increased from the physical value to $\frac{1}{2}\,m_s$ is of 
the expected size, comparable to the difference between $F_K$ and $F_\pi$.
The curvature makes it evident
that a linear extrapolation from values of order $\hat{m}\sim 
\frac{1}{2}\,m_s$ down to the physical region is meaningless. 

\section{$\mathbf{\pi\pi}$ scattering}
The experimental test of the hypothesis that the quark
condensate represents the leading order parameter relies on the
fact that $\lvac\, \qbar q\rvac$ not only manifests itself in the dependence of
the pion mass on $m_u$ and $m_d$, but also in the low energy properties of the 
$\pi\pi$ scattering amplitude.

At low energies, the scattering amplitude is dominated by the contributions
from the $S$-- and $P$--waves, because the angular momentum barrier suppresses
the higher partial waves. Bose statistics implies that configurations with
two pions and $\ell=0$ are symmetric in flavour space and thus 
carry either isospin $I=0$ or $I=2$, so that there are
two distinct $S$--waves. For $\ell=1$, on the other hand, the configuration
must be antisymmetric in flavour space, so that there is a single $P$--wave,
$I=1$. If the relative momentum tends to zero, only the $S$--waves contribute,
through the corresponding scattering lengths $a_0^0$ and $a_0^2$ (the
lower index refers to angular momentum, the upper one to isospin).

As shown by Roy \cite{Roy}, analyticity, unitarity and
crossing symmetry subject the partial waves to a set of coupled integral
equations. These equations involve two subtraction constants, which may be
identified with the two $S$--wave scattering lengths $a_0^0$, $a_0^2$. 
If these two constants are given, the Roy equations allow us to calculate the
scattering 
amplitude in terms of the imaginary parts above 800 MeV and the available
experimental information suffices to evaluate the relevant dispersion
integrals, to within small uncertainties \cite{ACGL}. In this sense, 
$a_0^0$, $a_0^2$ represent the essential parameters in low energy
$\pi\pi$ scattering. 

\begin{figure}[t] \centering

\psfrag{a0}{$a_0^0$}
\psfrag{a2}{$a_0^2$}
\includegraphics[width=5cm,angle=-90]{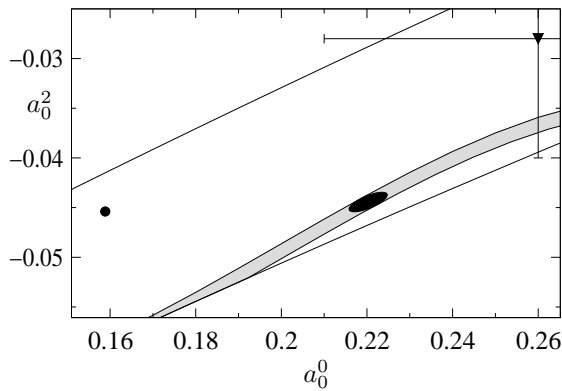}
\caption{\label{fig:aellipse} $S$--wave scattering
  lengths. The Roy
  equations only admit solutions in the ``universal band'',  
  spanned by the two tilted lines. The dot indicates Weinberg's leading order
  result, while the small ellipse includes the higher order corrections,
  evaluated in the framework of standard chiral perturbation theory. 
  In the generalized scenario, there is no prediction for $a_0^0$, but there
  is a correlation 
  between $a_0^0$ and $a_0^2$, shown as a narrow strip. The 
  triangle with error bars indicates the phenomenological
  range permitted by the old data, $a_0^0=0.26\pm 0.05$,
  $a_0^2=-0.028\pm0.012$ \protect\cite{Froggatt:1977hu}. }
\end{figure}

As a general consequence of the hidden symmetry, Goldstone bosons
of zero momentum cannot interact with one another. Hence the
scattering lengths $a_0^0$ and $a_0^2$ must vanish in the symmetry limit,
$m_u,m_d\rightarrow 0$. These quantities thus also measure the explicit
symmetry breaking generated by the quark masses, like $M_\pi^2$. In fact, 
Weinberg's low energy theorem~\cite{Weinberg 1966} states that, 
to leading order of the expansion in powers of $m_u$ and $m_d$,
the scattering lengths are proportional to $M_\pi^2$, the factor of
proportionality being fixed by the pion decay constant:\footnote{The
standard definition of the scattering length 
corresponds to $a_0/M_\pi$. It is not suitable in the present context, 
because it differs from the invariant 
scattering amplitude at threshold by a kinematic factor that diverges in the
chiral limit.}
\bea\label{eq:Weinberg} a_{0}^0=\frac{7 M_\pi^2}{32 \,\pi \, F_\pi^2}+
O(\hat{m}^2)\co
\hspace{1.3em}
a_{0}^2=-\frac{M_\pi^2}{16 \,\pi \,
  F_\pi^2}+O(\hat{m}^2)\fs\eea
Chiral symmetry thus provides the missing element: 
in view of the Roy equations,
Weinberg's low energy theorem fully determines the low energy behaviour of the
$\pi\pi$ scattering amplitude. The prediction (\ref{eq:Weinberg}) corresponds 
to the dot on the left of fig.~\ref{fig:aellipse}.

The prediction is of limited accuracy, 
because it only holds to leading order of the expansion in powers of the quark
masses. In the meantime, the chiral perturbation series of the scattering
amplitude has been worked out to two loops \cite{BCEGS}.
At first nonleading order of the expansion in powers
of momenta and quark masses,  the scattering amplitude
can be expressed in terms of $F_\pi$, $M_\pi$ and the 
coupling constants $\ell_1,\ldots\,,\ell_4$ that occur in the derivative
expansion of the effective Lagrangian at order $p^4$. The terms $\ell_1$ and
$\ell_2$ manifest themselves in the energy dependence 
of the scattering amplitude and can thus be determined phenomenologically. 
As discussed in section 5, the coupling constant $\ell_4$ is known rather
accurately from the dispersive analysis of the scalar form factor.
The crucial term is $\ell_3$ -- the range considered for this
coupling constant makes the difference between standard and generalized chiral
perturbation theory. In the standard framework, where the relevant scale
is in the range (\ref{eq:Lambda3}), one finds that the 
leading order result is shifted into the small ellipse shown
in fig.~\ref{fig:aellipse}, which corresponds to \cite{ABT,CGL}: 
\bea\label{eq:a0a2} 
a_0^0=0.220\pm0.005\co\hspace{2em}a_0^2=-0.0444\pm0.0010\fs\eea 
The numerical value quoted includes the higher order corrections 
(in the standard framework, the contributions
from the corresponding coupling constants are tiny).

The corrections from the higher order terms in the Gell-Mann-Oakes-Renner
relation can only be large if the estimate (\ref{eq:Lambda3}) for 
$\Lambda_3$ is totally wrong. As pointed out long ago \cite{GL 1983}, 
there is a low energy theorem
that holds to first nonleading order and relates the $S$--wave scattering
lengths to the scalar radius: 
\bea\label{eq:one loop}2a_0^0-5a_0^2=
\frac{3\,M_\pi^2}{4\pi F_\pi^2}\left\{1+
\frac{1}{3}\,M_\pi^2 \rs+\frac{41\,M_\pi^2}{192 \,\pi^2 F_\pi^2}\right\}
 + 
O(\hat{m}^3)\fs\eea
In this particular combination of scattering lengths, the term $\ell_3$ drops
out.  The theorem thus correlates the two scattering lengths, independently
of the numerical value of $\Lambda_3$. The correlation holds 
both in standard and generalized chiral perturbation theory. The corrections
occurring in eq.~(\ref{eq:one loop}) 
at order $\hat{m}^3$  have also been worked out. These are responsible for the
fact that the narrow strip, which represents the correlation in 
fig.~\ref{fig:aellipse}, is slightly 
curved.

\section{Impact of the new $\mathbf{K}$ decay data}
The final state interaction theorem implies that the phases of the form 
factors relevant for the decay $K\rightarrow \pi\pi e\nu$ are determined by
those of the $I=0$ $S$--wave and of the $P$--wave of elastic 
$\pi\pi$ scattering, respectively.
Conversely, the analysis of the final state distribution observed in this 
decay yields a measurement of the phase difference
$\delta(s)\equiv\delta_0^0(s)-\delta_1^1(s)$, in the region 
$4M_\pi^2<s<M_K^2$. As discussed above, the Roy equations determine
the behaviour of the phase shifts in terms of the two
$S$--wave scattering lengths. Moreover, in view of the correlation between
the two scattering lengths, $a_0^2$ is determined by $a_0^0$, so that the
phase difference $\delta(s)$ can be calculated as a function of $a_0^0$ 
and $q$, where $q$ is the c.m.~momentum
in units of $M_\pi$, $s= 4M_\pi^2(1+q^2)$.
In the region of
interest ($q<1$, $0.18<a_0^0<0.26$), the prediction reads 
\bea\label{eq:delta(s)} &&\delta_0^0-\delta_1^1=
\frac{q}{\sqrt{1+q^2}}\,( a_0^0+q^2\,b+q^4\,c+q^6\,d)\pm e\\
&&b= 0.2527+0.151\,\Delta a_0^0+1.14\,(\Delta a_0^0)^2 +
35.5\,(\Delta a_0^0)^3\co\no
&&c=0.0063-0.145\,\Delta a_0^0\co\hspace{2em}
d=-0.0096\co\nonumber\eea
with $\Delta a_0^0=a_0^0-0.22$.
The uncertainty in this relation mainly stems from the experimental input used
in the Roy equations and is not sensitive to $a_0^0$:
\bea\label{eq:errordelta} e= 0.0035 \,q^3+0.0015\,q^5\fs\eea
\begin{figure}[thb]
\leavevmode

\centering
\includegraphics[width=8cm]{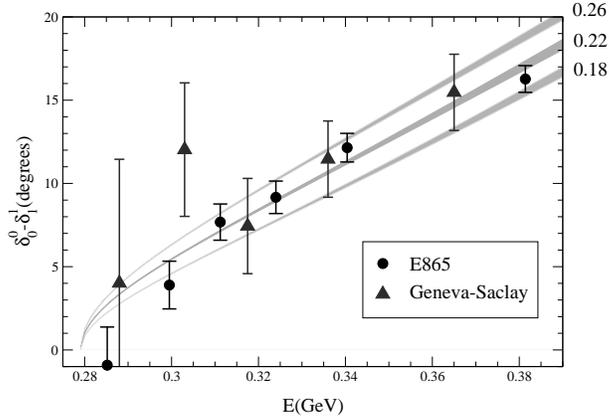}
\caption{\label{fig:deltaKl4} Phase relevant for the decay $K\rightarrow
  \pi\pi e\nu$. The three bands correspond to the 
three indicated values of the $S$--wave scattering length $a_0^0$. The
uncertainties are dominated by those from the experimental input used
in the Roy equations. The triangles are the data points of
ref.~\protect\cite{rosselet}, while the full circles represent the 
E865 results \protect\cite{Pislak}.}
\end{figure}

The prediction (\ref{eq:delta(s)}) is illustrated
in fig.~\ref{fig:deltaKl4}, where the energy dependence of  
the phase difference is shown for $a_0^0=0.18$, $0.22$ and
$0.26$. The width of the corresponding bands indicates the uncertainties,
which according to (\ref{eq:errordelta})
grow in proportion to $q^3$ -- in the range shown, they
amount to less than a third of a degree. 

The figure shows that the data of ref.~\cite{rosselet} barely 
distinguish between the three values of $a_0^0$ shown.  
The results of the E865 experiment at Brookhaven \cite{Pislak} are
significantly more precise, however. The best fit to these data is obtained
for $a_0^0=0.218$, with $\chi^2= 5.7$ for 5 degrees of freedom. This
beautifully confirms the value in eq.~(\ref{eq:a0a2}), 
obtained on the basis of standard chiral perturbation theory. 
There is a marginal
problem only with the bin of lowest energy: the corresponding scattering
lengths are outside the region
where the Roy equations admit solutions. In view of the
experimental uncertainties attached to that point, this discrepancy is without
significance: the difference between the central experimental value and 
the prediction amounts to 1.5 
standard deviations. Note also that the old data are perfectly consistent with
the new ones: the overall fit
yields $a_0^0=0.221$ with $\chi^2= 8.3$ for 10 degrees of freedom.

\begin{figure}[t]
\leavevmode
\centering
\includegraphics[width=8cm]{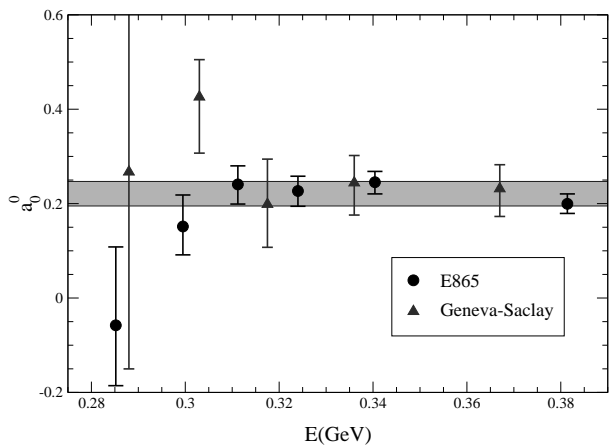}
\caption{\label{fig:aKe4}  $K_{e_4}$ data on the scattering length $a_0^0$.  
The triangles are the data points of
ref.~\protect\cite{rosselet}, while the full circles represent the 
E865 results \protect\cite{Pislak}. The horizontal band indicates the 
statistical average of the 11 values for $a_0^0$ shown in the figure.}
\end{figure}

The relation  (\ref{eq:delta(s)}) can be inverted, so
that each one of the values found for the phase difference yields
a measurement of the scattering length $a_0^0$.  
The result is shown in fig.~\ref{fig:aKe4}.
The experimental errors are remarkably small. It is not unproblematic,
however, to treat the data collected in the 
different bins as statistically independent: in the presence of correlations,
this procedure underestimates the actual uncertainties. Also, since the phase
difference rapidly rises with the energy, the binning procedure may introduce
further uncertainties. To account for this, the final result given in
ref.~\cite{CGLPRL},
\bea\label{eq:final result} a_0^0=0.221\pm 0.026\co\eea
corresponds to the 95\% confidence limit -- in effect, this amounts to
stretching the statistical error bar by a factor of two.

We may translate the result into an estimate for the magnitude 
of the coupling constant $\lbar_3$: the range (\ref{eq:final result})  
corresponds to $|\lbar_3|\,\lsim\, 16$.
Although this is a coarse estimate, it implies that the
Gell-Mann-Oakes-Renner relation does represent a decent approximation: 
 more than 94\% of the pion mass stems from the first term 
 in the quark mass expansion (\ref{eq:Mexp}), 
 i.e.~from the term that originates in the quark condensate.
This demonstrates that there is no need for a reordering of the chiral
perturbation series based on SU(2)$_{\indR}\times$SU(2)$_{\indL}$. In that
context, the generalized scenario has served its purpose and can now be
dismissed.    

A beautiful experiment is under way at CERN \cite{Nemenov}, 
which exploits the fact that $\pi^+\pi^-$ atoms decay into a pair of neutral
pions, through the strong transition
$\pi^+\pi^-\!\rightarrow\!\pi^0\pi^0$. Since 
the momentum transfer nearly vanishes, only the scattering lengths are
relevant: at leading order in isospin breaking, the transition amplitude is 
proportional to $a_0^0\!-\!a_0^2$. The corrections at
next--to--leading order are now also known~\cite{GLR}. Hence 
a measurement of the
lifetime of a $\pi^+\pi^-$ atom amounts to a measurement of
this combination of scattering lengths. At the planned accuracy of 10\% for
the lifetime, the experiment will yield a measurement of the scattering 
lengths to 5\%, thereby subjecting chiral perturbation theory to a very
sensitive test.

\section*{Acknowledgments}
It is a pleasure to thank Ji\v{r}i Ho\v{s}ek for a very pleasant
stay in Prague. Also, I thank the Institute for Nuclear Theory at the 
University of Washington for its hospitality and the Department of Energy for 
partial support during my visit to Seattle, where this report was written.

\end{document}